\documentclass[12pt]{JHEP3}
\usepackage{amsmath,amsfonts,amssymb,dsfont,ifthen,epsfig,slashed, cancel}

\newcommand{\be}{\begin{equation}}
\newcommand{\ee}{\end{equation}}
\newcommand{\ba}{\begin{eqnarray}}
\newcommand{\ea}{\end{eqnarray}}
\newcommand{\bas}{\begin{eqnarray*}}
\newcommand{\eas}{\end{eqnarray*}}
\newcommand{\bc}{\begin{center}}
\newcommand{\ec}{\end{center}}

\newcommand{\comment}[1]{}

\newboolean{Notes}\setboolean{Notes}{true}
\newcommand{\notes}[1]
            {\ifthenelse{\boolean{Notes}}{{\tt #1}}{}}

\newcommand{\csch}{\mathop{\rm csch\,}}
\newcommand{\sech}{\mathop{\rm sech\,}}

\preprint{arXiv:0712.4269 [hep-th]}

\title{Gravity Dual to Pure Confining Gauge Theory}
\author{Girma Hailu\thanks{hailu@lepp.cornell.edu}
\\
Newman Laboratory for Elementary Particle Physics \\
Cornell University\\
Ithaca, NY 14853}

\abstract{\\We find a dual gravity theory to pure confining $\mathcal{N}=1$ supersymmetric $SU(N)$ gauge theory in four dimensions which has the correct gauge coupling running in addition to reproducing the appropriate pattern of chiral symmetry breaking. It is constructed in type IIB string theory on $R^{1,3} \times R^1\times S^2\times S^3$ background with $N$ number of electric D5 and $2N$ number of magnetic D7-branes filling four dimensional spacetime and wrapping respectively two and four cycles.

}

\begin{document}

\emph{Introduction.---}The theory of quantum chromodynamics (QCD) of the strong nuclear interactions becomes highly nonperturbative and hard at low energies. The gauge/gravity duality \cite{Maldacena:1998re,Gubser:1998bc,Witten:1998qj} relates a gauge theory in strongly coupled nonperturbative region to a gravity theory in weakly coupled perturbative region and, therefore, provides the possibility for a calculable classical gravity description to low energy QCD. The first example of gauge/gravity duality in \cite{Maldacena:1998re} involves conformal field theory with $\mathcal{N}=4$ supersymmetry. A gravity dual to pure $\mathcal{N}=1$ supersymmetric $SU(N)$ gauge theory is highly desirable for several reasons. First, $\mathcal{N}=1$ supersymmetric $SU(N)$ gauge theory exhibits phenomena such as confinement and chiral supersymmetry breaking  and could serve as a laboratory to gain new insight into QCD. Second, there is a possibility that $\mathcal{N}=1$ supersymmetry may be part of nature at energies accessible in the coming generation of experiments at the Large Hadron Collider (LHC) and a gravity description is useful for calculating physical quantities such as glueball mass spectra in the $\mathcal{N}=1$ gauge theory. Third, if a suitable supersymmetry breaking scheme which removes the gaugino in the pure $\mathcal{N}=1$ theory is found, it could be used to study the real world QCD itself.
Fourth, the supergravity background has nonsingular geometry due to nonperturbative quantum effects which is useful for studying early universe cosmological scenarios and possibilities that a universe like ours could reside on the background.

Indeed, there has been extensive effort towards finding a gravity dual to pure $\mathcal{N}=1$ supersymmetric gauge theory, \cite{Polchinski:2000uf, Klebanov:2000hb, Maldacena:2000yy}  most notably laid the foundational work.  The work in \cite{Klebanov:2000hb} produced a gravity dual to $\mathcal{N}=1$ supersymmetric $SU(N+M)\times SU(N)$ gauge theory with $N$ number of D3 and $M$ number of D5-branes on $AdS_5\times T^{1,1}$ conifold background involving novel cascading renormalization group flow towards pure $\mathcal{N}=1$ supersymmetric $SU(M)$ gauge theory in the infrared and deformation of the conifold, but the gravity theory has constant dilaton and does not reproduce the gauge coupling running of a pure confining gauge theory.
The work in \cite{Maldacena:2000yy} produced supergravity solutions with $N$ number of NS5 or D5-branes involving running dilaton and appropriate pattern of chiral symmetry breaking, but it does not reproduce the gauge coupling running of pure $\mathcal{N}=1$ supersymmetric $SU(N)$ gauge theory in four dimensions.

In this note, we find a dual gravity theory to pure $\mathcal{N}=1$ supersymmetric $SU(N)$ gauge theory in four dimensions which has the correct gauge coupling running and which reproduces the appropriate pattern of chiral symmetry breaking robustly. 
The important new ingredients which facilitate our construction are
the gauge/gravity duality mapping with running dilaton and running axion obtained recently in \cite{Hailu:2007tm} with magnetic D7 and Dirac 8-branes playing crucial role and
the set of equations obtained in \cite{Hailu:2007ae} which allows studying systematically type IIB flows with $\mathcal{N}=1$ supersymmetry.

Our starting point is the running of the Yang-Mills coupling in the gauge theory and its mapping to the running of the dilaton in the gravity theory. We find a supergravity dual in type IIB string theory on $R^{1,3} \times R^1\times S^2\times S^3$  background with $N$ number of D5 and $2N$ number of D7-branes filling four dimensional (4-d) spacetime and wrapping respectively 2 and 4-cycles. The gauge theory is engineered by wrapping $N$ electrically charged D5-branes on non-zero $S^2$ cycle with $S^3$ of zero-size in the base at the tip which leads to blown-down $S^2$ and R-R 3-form $F_3$ flux through blown-up $S^3$ after the familiar geometric transition which deforms the tip as in \cite{Klebanov:2000hb}. See  \cite{Gopakumar:1998ki} for a conifold transition on a setting in the topological A-model with $S^2$ and $S^3$ interchanging roles.
The running of the gauge coupling leads to a running dilaton on the gravity side which is related to R-R $F_1$ flux such that the background follows the equations for the class of flows with imaginary self-dual 3-form flux in \cite{Hailu:2007ae} and the supergravity solutions are read off from \cite{Hailu:2007tm} with appropriate changes of variables. The runnings of the dilaton and the axion are due to magnetic coupling of the axion to D7-branes and Dirac 8-branes which emanate from the D7-branes.
Demanding that the correct renormalization group flow of the gauge theory living on the $N$ electrically charged D5-branes be reproduced leads to $2N$ magnetic D7-branes filling 4-d spacetime and wrapping 4-cycles at the ultraviolet edge of the background.
The background with the $F_1$ and the $F_3$ fluxes induces 3-from NS-NS $H_3$ flux and also 5-form R-R $F_5$ flux which can be viewed as coming from the wrapped D5-branes, which are fractional D3-branes, via backreaction NS-NS 2-form potential and the $F_3$ flux.
The axion potential $C_0$ in the axion-dilaton coupling coefficient is related to the Yang-Mills angle and preserves only a $Z_{2N}$ discrete symmetry in the ultraviolet which matches with the anomaly-free $R$-symmetry in the gauge theory. The supergravity solutions preserve only a $Z_2$ symmetry in the infrared and the breaking of the $Z_{2N}$ symmetry down to $Z_2$ gives $N$ discrete vacua, reproducing the same pattern of symmetry breaking by gaugino condensation in the gauge theory.

\emph{Gauge theory.---}Consider $\mathcal{N}=1$ supersymmetric pure $SU(N)$ gauge theory.
The classical theory has global $U(1)$ R-symmetry which is anomalous in the quantum theory. The anomaly-free quantum theory has a reduced $Z_{2N}$ discrete symmetry. Gaugino condensation breaks the $Z_{2N}$ symmetry down to $Z_2$ giving $N$ number of discrete vacua.
The low energy infrared dynamic of this theory at the scale $\Lambda$ is described by the Veneziano-Yankielowicz
superpotential \cite{Veneziano:1982ah},
\begin{equation}
W_{\mathrm{VY}}=NS-NS\log(\frac{S}{\Lambda^{3}}),\quad S=-\frac{1}{32\pi^2}\mathrm{Tr\,}\mathcal{W}^{\alpha}\mathcal{W}_{\alpha},\label{eq:rev4-5}
\end{equation}
where $S$ is the glueball superfield defined in terms of the gauge
chiral superfield $\mathcal{W}_{\alpha}$ containing the gauge and the gaugino fields in the $\mathcal{N}=1$ vector multiplet. Extremizing $W_{\mathrm{VY}}$ with $S$ gives the vacuum expectation value of the glueball superfield corresponding to the $N$ vacua,
\be
\langle S \rangle= \Lambda^{3} e^{2\pi i k/N},\quad  k=1, 2,\cdots,N.
\ee
Let us define $T={8\pi^2}/{g^2}$, where $g$ is the Yang-Mills coupling constant in the gauge theory. The quantum loop corrections to the running of the gauge coupling are exhausted at one loop and with the exact $\beta$ function we have
\be
\frac{d T}{d\ln(\Lambda/\Lambda_c)}=3N,\label{dTdmu-1}
\ee
where we will take $\Lambda_c$ to be the scale at the infrared end of the theory.

\emph{Dilaton.---}The gauge coupling is related to the string coupling such that
$T={2\pi e^{-\Phi}}/{g_s}$,
where $g_s$ is the bare string coupling and $\Phi$ is the dilaton.
The scale of the gauge theory is mapped to the radius $r$ such that
${\Lambda}/{\Lambda_c}={r}/{r_c}$, where $r_c$ is the smallest value of $r$ at the tip and $\ln(r/r_c)\ge 0$.
The appropriate running of the dilaton then follows,
\be
\frac{d}{d\tau}(e^{-\Phi})=\frac{g_s N}{2\pi},\quad \tau =3\ln(r/r_c).\label{dilaon-1}
\ee

\emph{Metric and fluxes.---}The supergravity background we have here is similar to the one studied in \cite{Hailu:2007tm, Hailu:2006uj} with no regular D3-branes and the running of the dilaton now given by (\ref{dilaon-1}). Therefore, it is accommodated in the class of supergravity flows with imaginary self-dual 3-form flux given in \cite{Hailu:2007ae} with running dilaton and running axion. The supergravity solutions can be read off from \cite{Hailu:2007tm}
with $-{s}/{(3+d)}$ in \cite{Hailu:2007tm} replaced by ${N}/{2\pi}$ and the radial variable $\tau$ redefined as in (\ref{dilaon-1}). The difference in the signs is because the supergravity flow we have here is dual to a confining gauge theory and the magnitude of the dilaton increases towards the infrared which is opposite to the behavior of the dilaton in \cite{Hailu:2007tm}.
The metric and flux ansatz used in \cite{Hailu:2007tm} was taken from \cite{Butti:2004pk, Papadopoulos:2000gj} with some modifications and additions to accommodate the runnings of the dilaton and the axion. We work in the string frame. For the metric,
\begin{equation}
ds^2=e^{2A(\tau)}\eta_{\mu \nu} dx^\mu dx^\nu + ds_6^2(y),\quad
ds_6^2(y)=\delta_{mn}G^m G^n,\label{10dmetric2}
\end{equation}
where $x^\mu$ are the coordinates on $R^{1,3}$ in 4-d, $y^m$ are the coordinates in the extra 6-d space, and $\tau$ is the radial variable defined in (\ref{dilaon-1}) and parameterizes $R^1$ in $R^{1,3} \times R^1\times S^2\times S^3$.
The $G^m$ are real differential 1-forms which are expressed in terms of linear combinations of the coordinate 1-forms $dy^n$ on $Y$ with coefficients which are functions of $y$,
\ba
G^1=e^{\frac{x+g}{2}}e_1,\quad G^2=\mathcal{A} \,e^{\frac{x+g}{2}}e_2+\mathcal{B} \,e^{\frac{x-g}{2}}\tilde{\epsilon_2},\quad G^3=e^{\frac{x-g}{2}}\tilde{\epsilon_1},\nonumber \\G^4=\mathcal{B} \,e^{\frac{x+g}{2}}e_2-\mathcal{A} \, e^{\frac{x-g}{2}}\tilde{\epsilon_2},\quad G^5=e^{\frac{-6p-x}{2}}d\tau,\quad G^6=e^{\frac{-6p-x}{2}}\tilde{\epsilon_3},\label{Gm-defn}
\ea
\begin{eqnarray}&
e_1=d\theta_1,\quad e_2=-\sin{\theta_1}d\phi_1,\quad
\tilde{\epsilon_1}=\epsilon_1-a e_1,\quad
\tilde{\epsilon_2}=\epsilon_2-a e_2, &\nonumber\\&
\tilde{\epsilon_3}=\epsilon_3+\cos \theta_1 d\phi_1,\quad
\epsilon_1=\sin \psi \sin\theta_2 d\phi_2+\cos\psi d\theta_2,
&\nonumber\\& \epsilon_2=\cos \psi \sin\theta_2 d\phi_2-\sin\psi
d\theta_2,\quad \epsilon_3=d\psi+\cos\theta_2
d\phi_2,&\label{epsiloni}
\end{eqnarray}
with $\theta_1,\,\theta_2\, \in [0,\pi]$, $\phi_1,\,\phi_2\, \in [0,2\pi]$, and $\psi\, \in [0,4\pi]$ parameterizing the base $S^2\times S^3$ in $R^{1,3} \times R^1\times S^2\times S^3$.
The NS-NS $H_3$ and R-R $F_3$, $F_5$, and $F_1$ fluxes take the form
\be
H_3=d\tau\wedge \Bigl(h_1'(\epsilon_1\wedge \epsilon_2+e_1\wedge
e_2) +h_2'(\epsilon_1\wedge e_2-\epsilon_2\wedge
e_1)\Bigl)+h_2\tilde{\epsilon}_3\wedge (\epsilon_1\wedge
e_1+\epsilon_2\wedge e_2),\label{fluxes}
\ee
\be
F_3=-\frac{1}{4}\alpha' N e^{-\Phi}\Bigl(\tilde{\epsilon}_3\wedge (\epsilon_1\wedge
\epsilon_2+e_1\wedge e_2-b (\epsilon_1\wedge e_2-\epsilon_2\wedge
e_1))+b' d\tau\wedge (\epsilon_1\wedge
e_1+\epsilon_2\wedge e_2)\Bigr),\label{F3-ansatz}
\ee
\be
F_5=K\,e_1\wedge e_2 \wedge \epsilon_1 \wedge \epsilon_2
\wedge \epsilon_3,
\quad
K=-\frac{1}{2}\alpha' N  e^{-\Phi}(h_1+bh_2), \quad F_{1}=-\frac{N}{2\pi}\,\tilde{\epsilon}_3.\label{khh1h2}
\ee
The variables $A$, $\mathcal{A}$, $\mathcal{B}$, $a$ $x$, $g$, $p$, $h_1$, $h_2$, $b$, and $K$   in the metric and in the fluxes above are functions of $\tau$ with $\mathcal{A}^2+\mathcal{B}^2=1$. The factor of $e^{-\Phi}$ in the ansatz for $F_3$ is put in order to obtain $dF_3=-F_1\wedge H_3$.

\emph{Solutions.---}Now we want to read off the supergravity solutions for all the variables in the metric and in the fluxes from \cite{Hailu:2007tm} with the parameter ${s}/{(3+d)}$ replaced by $-{N}/{2\pi}$.
First, the running of the dilaton is given by
\be
e^\Phi=\frac{1}{e^{-\Phi_c}+\frac{g_sN}{2\pi}\tau},\label{phi-soln}
\ee
where $\Phi_c$ is the value of the dilaton at $\tau=0$.
The solutions involve the relations
\be
a=-\sech \tau,\quad b=-\tau \csch \tau,
\quad
e^{g}=\tanh \tau,\quad
\mathcal{A}=\tanh \tau,\quad \mathcal{B}=\sech \tau,\label{Us0b}
\ee
\be
h_2=\frac{1}{4}\alpha'g_s N(1-\tau \coth \tau)\csch \tau,\,\,
h_{1}=h_{2} \cosh \tau,\,\, K=-\frac{1}{2}e^{-\Phi}\alpha' N (h_1+bh_2).\label{h1h2K-soln1}
\ee
The expressions (\ref{Us0b})-(\ref{h1h2K-soln1}) are read off from \cite{Butti:2004pk} and are the same as in the solutions in \cite{Klebanov:2000hb}  written in the form given in \cite{Butti:2004pk} with nonzero $\Phi$.
The $F_1$ flux is
\be
F_1=-\frac{N}{2\pi} e^{\frac{6p+x}{2}}G^6=-\frac{N}{2\pi}\tilde{\epsilon}_3.\label{F1-tot1}
\ee
Defining $v=e^{6p+2x}$, $u=e^{2x}$ and $h=e^{-4A}$, the equations for $v$, $u$ and $h$ are
\be
v'+(2\coth \tau+\frac{3g_sN}{2\pi}  e^\Phi)v-3=0,
\label{v-eq}
\ee
\be
\frac{d}{d\tau}\ln(\frac{u}{h}) =\frac{2}{v}-\frac{g_sN}{\pi}  e^\Phi, \label{uh-soln}\ee
\be h'+\frac{1}{4}g_s e^\Phi K \frac{h}{u}-\frac{g_sN}{2\pi}  e^\Phi h=0,\label{h-soln}
\ee
with $\Phi$ given by (\ref{phi-soln}) and $K$ given in (\ref{h1h2K-soln1}).
A consistent set of solutions requires the boundary conditions $v(0)=0$, $u(0)=0$, and $h(0)=h_0$. Equation (\ref{v-eq}) is easily solved for $v$, (\ref{uh-soln}) is then solved for $u/h$ and the result is used for $h/u$ in (\ref{h-soln}) to write integral solution for $h$ which is then used with the expression for $u/h$ to find $u$. The values of $u$ and $v$ increase towards the ultraviolet as $\tau$ increases. The value of $h$ also starts increasing as $\tau$ increases from $\tau=0$ because of the sign of the third term in (\ref{h-soln}) which comes from the running of the dilaton in the asymptotically-free gauge theory here and is opposite to that in \cite{Hailu:2007tm} where $h$ decreases.

\emph{D7 and Dirac 8-branes.---}As it is shown in \cite{Hailu:2007tm}, the runnings of the axion and the dilaton are due to magnetic D7-branes filling 4-d spacetime and wrapping the 4-cycle
\be
\omega_4=\sin \theta_1 \sin \theta_2 \, d\theta_1 \wedge d\phi_1 \wedge d\theta_2 \wedge d\phi_2
=\frac{1}{u}G^1\wedge G^2\wedge G^3\wedge G^4.
\ee
The electric dual to a D7-brane is D(-1)-brane which is a point in spacetime.
The magnetic D7-branes give the $F_1$ flux through $\tilde{\epsilon}_3$ and the $F_1$ flux is not closed because of Dirac 8-branes which emanate from the D7-branes, in the same way a Dirac string emanating from a magnetic monopole in 4-d carries magnetic flux along the string.
In particular, it is shown in \cite{Hailu:2007tm} that the $F_1$ flux satisfies
\be F_1-\star \bar{\mathcal{F}}_9=dC_0,\label{F1-tot4}
\ee
where $\star \bar{\mathcal{F}}_9$ accounts for the flux due to the Dirac 8-branes which makes $dF_1\ne0$. The magnetic flux $\mathcal{F}_9=dC_8=-\star F_1$, where $C_8$ is the D7-branes potential, is closed as a result of contributions to the flux from the magnetic charge of the D7-branes and from the Dirac 8-branes, $\mathcal{F}_9=-\star dC_0+\bar{\mathcal{F}}_9$. The axion $C_0$ and the magnitude of the magnetic D7-branes charge $\tilde{Q}_7$ are obtained using the expression for $C_0$ given in \cite{Hailu:2007tm} with ${s}/{(3+d)}\to-{N}/{2\pi}$,
\be
C_0=-\frac{N}{2\pi}\psi\,,\quad \tilde{Q}_7=\int_{4\pi}^0 dC_0= 2N.\label{C0-soln}
\ee
Therefore, we have $2N$ number of D7-branes.

\emph{Bottom of background.---}Now we explore the geometry at the bottom of the background. Let us expand the variables in the metric given by (\ref{Us0b}) and solving (\ref{v-eq})-(\ref{h-soln}) to leading order in $\tau$ near $\tau= 0$,
\ba
e^g= \tau+\mathcal{O}(\tau^3), \quad a=-1+\mathcal{O}(\tau^3),\quad \mathcal{A}=\tau+\mathcal{O}(\tau^3),\quad \mathcal{B}=1+\mathcal{O}(\tau^2),\nonumber\\
h=h_0+\mathcal{O}(\tau),\quad v= \tau+\mathcal{O}(\tau^2),\quad u= e^{2\Phi_c} h_0 \tau^2+\mathcal{O}(\tau^3),
\ea
where $h_0=h(0)$.
Therefore, near $\tau=0$, we have $e^{x+g}={u^{1/2}}{e^{g}}= e^{\Phi_c} h_0^{1/2}\tau^2+\mathcal{O}(\tau^3)$, $e^{x-g}={u^{1/2}}/{e^{g}}= e^{\Phi_c} h_0^{1/2}+\mathcal{O}(\tau)$, $e^{-6p-x}={u^{1/2}}/{v}= e^{\Phi_c} h_0^{1/2}+\mathcal{O}(\tau)$. The 1-forms in the metric given by (\ref{10dmetric2}) near $\tau=0$ have the forms
$G^1\sim  (e^{2\Phi_c} h_0)^{1/4}\tau e_1$,  $G^2\sim  (e^{2\Phi_c} h_0)^{1/4}\tilde{\epsilon}_2$, $G^3\sim  (e^{2\Phi_c} h_0)^{1/4}\tilde{\epsilon}_1$, $G^4\sim -(e^{2\Phi_c} h_0)^{1/4} \tau\epsilon_2$,  $G^5\sim  (e^{2\Phi_c} h_0)^{1/4} d\tau$, and $G^6\sim  (e^{2\Phi_c} h_0)^{1/4}\tilde{\epsilon}_3$.
The metric near $\tau=0$ is then
\be
ds^2\sim h_0^{-1/2} \eta_{\mu \nu} dx^\mu dx^\nu
+ e^{\Phi_c} h_0^{1/2}\left( d\tau^2+ \tilde{\epsilon}_1^2+\tilde{\epsilon}_2^2+\tilde{\epsilon}_3^2
+\tau^2(e_1^2+\epsilon_2^2)\right),\label{ds2-tip}
\ee
where the $\tilde{\epsilon}_i$ are given by (\ref{epsiloni}) with $a\sim-1$.
Therefore, the geometry at $\tau=0$ is
$R^{1,3}\times S^3$ with the radius of $S^3$ a function of the magnitude of the vacuum expectation value of the glueball superfield $S$ at $\tau=0$ which is simply the 't Hooft coupling \cite{'tHooft:1973jz} $e^{\Phi_c}g_s N$ and confinement via gaugino condensation is the source for the deformation of the tip as established in \cite{Klebanov:2000hb}. The boundary value $h_0$ in solving (\ref{v-eq})-(\ref{h-soln}) is then related to the radius of $S^3$ at $\tau=0$ and, therefore, a function of the 't Hooft coupling. 

\emph{Large $N$ and large 't Hooft coupling.---}Let us check the conditions on the parameters in the theory for the supergravity description to be good. We need a large value of $N$ so that only planar Feynman graphs survive and a large 't Hooft coupling. The 't Hooft coupling is related to the magnitude of the vacuum expectation value of the glueball superfield  and the size of the holes in 't Hooft's ribbon graphs have large size and can define Riemann surface for a string worldsheet for large 't Hooft coupling if filled by D-brane disks \cite{Ooguri:2002gx}. On the other hand, for small 't Hooft coupling, the gauge theory has a good perturbative description.
Let us see the implication of the large 't Hooft coupling constraint in our case, particulary now that we have a dilaton whose magnitude decreases as $\tau$ increases. The running of the dilaton is given by (\ref{phi-soln}). In order for $e^{\Phi}g_sN>>1$, we need $e^{\Phi_c}g_sN(1-\frac{\tau}{2\pi})>>1$. Therefore, the region in which the supergravity description is good is bounded by $r/r_c <e^{2\pi/3}$ and presents a robust region for large $e^{\Phi_c}g_sN$. We need the gauge/gravity flow to be within $0\le \tau <2\pi$ and we take the ultraviolet cutoff of the gauge theory at the scale $\Lambda_{0}=\Lambda_c e^{2\pi/3}$ or, equivalently, the radius at the edge of the background is $r_{0}= r_c e^{2\pi/3}$. The background is noncompact and can be thought of as glued to a larger background at $r=r_{0}$.

\emph{Chiral symmetry breaking.---}Next let us see how the chiral symmetry in the gauge theory shows up in the gravity theory. The gauge theory at the classical level has global $U(1)$ R-symmetry which corresponds in the gravity theory to the symmetry $\psi\to\psi+c$, where $c$ is constant, as shown in \cite{Klebanov:2000hb, Maldacena:2000yy}. The anomaly-free quantum theory has only the $Z_{2N}$ discrete symmetry.
Now we have the $C_0$ potential which is nonzero in the ultraviolet and which depends on $\psi$ and we want to see that it reproduces the same symmetry.
The axion-dilaton coupling coefficient is given by
\be
\tau_{ad}=\frac{i}{g_s}e^{-\Phi}+C_0=\frac{i}{g_s}e^{-\Phi}-\frac{N}{2\pi}\psi,\label{tau-ad}
\ee
where we have used (\ref{C0-soln}) for $C_0$ and the subscript in $\tau_{ad}$ is for axion-dilaton in order to avoid confusion in notation with the radial variable $\tau$.  Note that the second term in (\ref{tau-ad}) corresponds to a Yang-Mills angle of $\Theta=-N\psi$ and clearly shows that $\psi\to\psi+c$, where $\psi=\psi+4\pi$, is anomalous $U(1)$ symmetry and $\Theta$ is left invariant under $\psi \to \psi+\frac{4\pi}{2N}n$ which gives $\Theta\to \Theta-2\pi n$.
Therefore, $\tau_{ad}$ preserves a reduced $Z_{2N}$ symmetry with $\psi \to \psi+\frac{4\pi}{2N}n$, where $1\le n\le 2N$. This discrete symmetry in $\tau_{ad}$ corresponds to the symmetry in the locations on $\psi$, where $d\psi$ is related to $G^6$ as given in (\ref{Gm-defn}) and (\ref{epsiloni}), at which the D7-branes in the ultraviolet edge wrap the $\omega_4$ cycle. Both the $F_1$ flux and the $F_3$ flux in the ultraviolet contain $d\psi$ and are single-valued.
The fluxes in the supergravity solutions have non-vanishing components which explicitly contain $\sin \psi$ and $\cos \psi$ in the infrared as we see in the expressions (\ref{fluxes}) and (\ref{F3-ansatz}) together with (\ref{Us0b}) and (\ref{h1h2K-soln1}) and preserve only a $Z_2$ symmetry corresponding to $\psi\to\psi+2\pi$. Therefore, the $Z_{2N}$ symmetry is broken  down to $Z_2$ by the solutions in the gravity theory and gives the same $N$ number of discrete vacua in the infrared as in the gauge theory.

\emph{Conclusions.---}The final picture we have is that the electric $\mathcal{N}=1$ supersymmetric $SU(N)$ gauge theory lives on the electrically charged $N$ number of D5-branes filling 4-d spacetime and wrapping $S^2$ with vanishing $S^3$ at the infrared end before geometric transition. The supergravity solutions involve the background with fluxes after the transition with $S^2$ blown-down and $S^3$ of finite size at the tip which gives the familiar gravitational description to confinement via gaugino condensation in the gauge theory. The $2N$ number of magnetic D7-branes fill up 4-d spacetime and wrap 4-cycles at the ultraviolet edge with invisible Dirac 8-branes filling 4-d spacetime and emanating from the D7-branes and the $F_1$ flux through $G^6$ which is related to the running of the dilaton. We also have the backreaction NS-NS $H_3$ flux and the $F_5$ flux effectively coming from the wrapped D5 fractional D3-branes.
The quantum theory has $Z_{2N}$ discrete symmetry in the ultraviolet and arises robustly from our solution for the axion potential. The $Z_{2N}$ symmetry is broken down to $Z_2$ by the supergravity solutions in the infrared giving $N$ vacua as in the gauge theory.
It is satisfying that the $C_0$ potential which comes from the $F_1$ flux which is a crucial component of our construction obtained using the equations for type IIB flows with $\mathcal{N}=1$ supersymmetry we obtained in \cite{Hailu:2007ae} and the gauge/gravity duality mapping with running dilaton and running axion we obtained in \cite{Hailu:2007tm} has provided a consistent picture. Most importantly, the renormalization group flow of the gauge theory is reproduced in the gravity theory.
What can we say about the gauge theory which lives on the $2N$ magnetic D7-branes? Because the number of D7-branes is the same as the order in the $Z_{2N}$ discrete symmetry in the ultraviolet, it is convenient to wrap each one of the D7-branes over a 4-cycle at each one point $\psi = \frac{4\pi}{2N}n$, $1\le n\le 2N$, in the ultraviolet edge of the background, and the total charge is the same. We then have $U(1)^{2N}$ gauge theory living on the D7-branes which is infrared-free and can be simply ignored.
The supergravity solutions we have presented could be used to study additional gravitational descriptions to features in the gauge theory and the metric could be used to calculate physical quantities such as glueball mass spectra. We end with a conjecture:

\emph{Type IIB string theory with N electric D5 and 2N magnetic D7-branes on $R^{1,3} \times R^1\times S^2\times S^3$ background is dual to pure $\mathcal{N}=1$ supersymmetric $SU(N)$ electric Yang-Mills theory in the large $N$ limit.}

\emph{Acknowledgements.---}
We are grateful to Henry Tye for helpful discussions.
This research is supported in part by the National Science Foundation under grant number
NSF-PHY/03-55005.

\newpage

\bibliographystyle{JHEP}

\providecommand{\href}[2]{#2}\begingroup\raggedright\endgroup

\end{document}